# An allometric scaling relation based on logistic growth of cities

Yanguang Chen

Department of Geography, College of Urban and Environmental Sciences, Peking University, Beijing 100871, China

**Abstract**: The relationships between urban area and population size have been empirically demonstrated to follow the scaling law of allometric growth. This allometric scaling is based on exponential growth of city size and can be termed "exponential allometry", which is associated with the concepts of fractals. However, both city population and urban area comply with the course of logistic growth rather than exponential growth. In this paper, I will present a new allometric scaling based on logistic growth to solve the abovementioned problem. The logistic growth is a process of replacement dynamics. Defining a pair of replacement quotients as new measurements, which are functions of urban area and population, we can derive an allometric scaling relation from the logistic processes of urban growth, which can be termed "logistic allometry". The exponential allometric relation between urban area and population is the approximate expression of the logistic allometric equation when the city size is not large enough. The proper range of the allometric scaling exponent value is reconsidered through the logistic process. Then, a medium-sized city of Henan Province, China, is employed as an example to validate the new allometric relation. The logistic allometry is helpful for further understanding the fractal property and self-organized process of urban evolution in the right perspective.

# 1 Introduction

Allometry is originally a concept of biology concerning the study of the relationship between size and shape. Allometric growth implies the increase in size of different organs or parts of an organism at various relative rates of growth (Damuth, 2001; He and Zhang, 2006; Small, 1996). The law of allometric growth was originally introduced to urban studies by Naroll and Bertalanffy (1956), modeling the relationships between urban and rural population. Then, it was employed to describe the relationship between a system of cities and the largest city within the urban system (Beckmann, 1958; Carroll, 1982). From then on, a large number of works were devoted to exploring the scaling relationships between urban area and population, indicating size and shape in the growth of human communities (Batty and Longley, 1994; Chen, 2010; Lee, 1972; Lee, 1989; Lo and Welch, 1977; Longley, 1991; Nordbeck, 1971; Tobler, 1969). The empirical studies put the allometric analyses of cities in a dilemma of dimension (Lee, 1989). Fortunately, the concept of fractal dimension raised the allometric modelling phoenix-like from the ashes. The allometry is always associated with fractal structure (Batty and Longley, 1994; Chen, 2010; Enguist *et al*, 1998; He, 2006; West, 2002; West *et al*, 1997), and can be applied to many fields of urban researches (Batty *et al*, 2008; Bettencourt *et al*, 2007; Chen and Jiang, 2009; Chen and Lin, 2009; Kühnert *et al*, 2006; Samaniego and Moses, 2008). Today, allometric scaling has become one of basic laws in urban geography (Lo, 2002), and, it will be more important in future urban studies (Batty, 2008).

The traditional allometry is in fact based on exponential growth (Bertalanffy, 1968). However, the urban growth in the real world is a logistic process rather than exponential growth. For example, the relationships between urban area and population are supposed to follow the law of allometric growth, which is based on geometric growth or exponential distribution. However, if we apply the allometric scaling law to the observational datasets of the time series of city size and urban area from the real world, the effects are often unsatisfactory and thus unconvincing. The cause is that, in many cases, the courses of urban growth take on sigmoid curves rather than exponential curves. For the geometric growth, the allometric scaling exponent is a constant, which is equal to the ratio of the relative rate of growth of one element (say, urban area) to that of another element (say, urban population). For the logistic growth, the scaling exponent does not equal the ratio of relative growth rates, which is not a constant. In this instance, we need a new allometric scaling relation based on logistic growth to model urban evolution.

Allometric scaling falls into two types: *longitudinal allometry* and *cross-sectional allometry*. The former is based on a process of a city's growth within a long period of time, while the latter is based on a distribution of a set of cities within a geographical region at a given time. A logistic process is a dynamical process of replacement (Chen, 2012; Rao *et al*, 1989). Based on logistic growth and basic measures such as city size, a new measure termed *replacement quotient* can be defined by the ratio of one basic measure to the difference between the measure and the capacity i.e., the maximum amount of the measure. Using this replacement quotient, I will propose a new allometric scaling relation on the grounds of logistic processes. Therefore, the longitudinal allometry can be divided into *exponential allometry* and *logistic allometry*. The former is based on geometric growth, while the latter is based on sigmoid growth.

This paper aims at researching the longitudinal allometry of urban growth. The remaining parts are arranged as follows. In Section 2, an allometric relation based on logistic growth is proposed. The model is simple and easily understood. In Section 3, an empirical case is presented by using the observational data. The least squares computation is employed to make a longitudinal allometric analysis of urban evolution of Xinyang, China. In Section 4, several related questions are discussed. Finally, the paper is concluded with a brief summary of this study. The significance of this work lies in four aspects: first, it provided a new model of longitudinal allometry for urban theoretical studies; second, it explained the allometric scaling exponent in its right perspective; third, it laid a foundation for the future allometric studies based on sigmoid growth rather than geometric growth; fourth, it associated the allometric scaling with the dynamical analysis of spatial replacement.

## 2 Mathematical models

### 2.1 Allometric growth based on exponential growth

Allometric growth means that different parts of a system have different relative rates of growth. If the system follows the law of allometric growth, the relationship between two parts can be described with a power function such as

$$y_t = ax_t^b, \tag{1}$$

where $t$ denotes time, $x_t$ is one measure of time $t$, $y_t$ is another measure of time $t$, which corresponds to $x_t$, two parameters, $a$ and $b$ refer to the proportionality coefficient and scaling exponent, respectively. Taking derivative of equation (1) with respect to *t* yields

$$b = \frac{dy_t/(y_t dt)}{dx_t/(x_t dt)} = \frac{dy_t/y_t}{dx_t/x_t}, \quad (2)$$

which suggests that the ratio of the relative rate of growth of one part to that of the other part is a constant. This is just the mathematical essence of the law of allometric growth (Chen and Jiang, 2009; West and Griffin, 1999).

An allometric growth equation can be theoretically derived from two equations of exponential growth (Bertalanffy, 1968; Chen and Jiang, 2009). In other words, equation (1) is based on two exponential functions in the forms

$$x_t = x_0 e^{r_x t}, \quad (3)$$

$$y_t = y_0 e^{r_y t}, \quad (4)$$

in which $x_0$ refers to the initial value of *x*, $y_0$ to the initial value of *y*, $r_x = dx_t/(x_t dt)$ refers to the relative rate of growth of *x*, and $r_y = dy_t/(y_t dt)$ to the relative rate of growth of *y*. In light of equation (2), we have a ratio $b = r_y/r_x$. For the scaling relation between urban area and population, it can be proved that the reasonable range of the *b* value comes between 2/3 and 1 (Chen, 2010).

The allometric scaling is in fact a proportional relation between different measures of a system. The precondition for one measurement being proportional to another measurement is that the two measurements have identical dimension. Suppose that the dimension of *x* (e.g., city population) is $D_x$, and that of *y* (e.g., urban area) is $D_y$. According to the principle of dimension consistency (Lee, 1989), equation (1) can be rewritten as

$$y_t = a x_t^{D_y/D_x}. \quad (5)$$

Thus we have a fractal parameter equation as below:

$$b = \frac{D_y}{D_x} = \frac{r_y}{r_x}, \quad (6)$$

which associates the space parameter, i.e., fractal dimension *D*, and the time parameter, i.e., the relative rate of growth, *r*. This indicates that the geographical process based on time and the geographical patterns indicative of space are the different sides of the same coin. Especially, the allometric scaling exponent is a function of fractal dimension of urban form (Batty and Longley, 1994; Chen, 2010).

## 2.2 Allometric growth based on logistic growth

An exponential growth suggests a process of expanding or increase in size without constraint and limitation. This is a kind of positive feedback, which leads to the final breakdown of a system. If the growth is confined by an upper limit (e.g., the maximum value), the exponential function will be replaced by a sigmoid function. That is, the exponential growth curve will be squashed into a sigmoid curve by the lower and upper limits (e.g., the initial value and the maximum value) (Chen, 2012). Among various sigmoid functions, the logistic function is the best one in character and familiar to scientists. For $x$ and $y$, the logistic models can be expressed as

$$x_t = \frac{x_{max}}{1+(x_{max}/x_0 -1)e^{-k_x t}}, \qquad (7)$$

$$y_t = \frac{y_{max}}{1+(y_{max}/y_0 -1)e^{-k_y t}}, \qquad (8)$$

where $x_{max}$ and $y_{max}$ are both the parameters of capacity, that is, the maximum values of $x$ and $y$, $x_0$ and $y_0$ are the initial values of $x$ and $y$, and $k_x$ and $k_y$ denote the original rates of growth of $x$ and $y$, respectively. A logistic growth curve takes on the shape of the letter S. The S-shaped curve is of odd symmetry. The first phase ($x_t<x_{max}/2$, $y_t<y_{max}/2$) is similar to exponential curve, while the second phase ($x_t>x_{max}/2$, $y_t>y_{max}/2$) is similar to logarithmic curve. If the capacity parameters are large enough, and the $t$ value is not large, it follows $1/x_{max} \to 0$, $1/y_{max} \to 0$, and we will have

$$x_t = \frac{1}{1/x_{max} + (1/x_0 - 1/x_{max})e^{-k_x t}} \approx x_0 e^{k_x t} \approx x_0 e^{r_x t}, \qquad (9)$$

$$y_t = \frac{1}{1/y_{max} + (1/y_0 - 1/y_{max})e^{-k_y t}} \approx y_0 e^{k_y t} \approx y_0 e^{r_y t}, \qquad (10)$$

in which $k_x \approx r_x$, $k_y \approx r_y$. That is, if $x_t<x_{max}/2$, $y_t<y_{max}/2$, equations (7) and (8) will be reduced to equations (3) and (4), from which it follows equation (1). This suggests that, for the early stage of logistic growth, there exists an approximate allometric scaling relationship between $x$ and $y$. however, this is a local allometric relation instead of a global allometric relation. For the whole process of logistic growth, we cannot infer an allometric relationship between the two measures, $x$ and $y$.

If we define a new measure termed *replacement quotient*, we can derive an allometric scaling relation from logistic equations of growth. Equations (7) and (8) can be transformed into the following forms:

$$\frac{x_t}{x_{\max} - x_t} = \frac{x_0}{x_{\max} - x_0} e^{k_x t}, \quad (11)$$

$$\frac{y_t}{y_{\max} - y_t} = \frac{y_0}{y_{\max} - y_0} e^{k_y t}, \quad (12)$$

From equations (11) and (12) it follows

$$\frac{y_t}{y_{\max} - y_t} / \frac{y_0}{y_{\max} - y_0} = \left(\frac{x_t}{x_{\max} - x_t} / \frac{x_0}{x_{\max} - x_0}\right)^{k_y / k_x}, \quad (13)$$

which suggests an allometric substitution. A process of logistic growth is in fact a process of replacement dynamics (Chen, 2012; Rao *et al*, 1989). A type of constituent elements is substituted by another type of elements, and a kind of activities is substituted with another kind of activities, and so on. For example, in a process of urbanization, rural population is replaced with urban population, and rural land is replaced by urban land. If $x_{\max}$ and $y_{\max}$ represent the maximum amounts which can be replaced, then $x_t$ and $y_t$ will represent the cumulative amounts of replacement at the time of $t$. Defining the replacement quotients of $x$ and $y$ as $X_t = x_t/(x_{\max} - x_t)$, $Y_t = y_t/(y_{\max} - y_t)$, we can rewrite equation (13) as an allometric relation:

$$Y_t = \eta X_t^\sigma, \quad (14)$$

where $X_0 = x_0/(x_{\max} - x_0)$ and $Y_0 = y_0/(y_{\max} - y_0)$ denotes the initial quotients of replacement, $\eta = Y_0 X_0^{-\sigma}$ is a proportionality coefficient, $\sigma = k_y/k_x$ is the allometric scaling exponent of replacement dynamics.

## 3 Empirical analysis

### 3.1 Measures and materials

In this paper, the regression analysis based on ordinary least square method is used to make a longitudinal allometric analysis of urban evolution of Xinyang, a medium-sized city of Henan Province, China. Two basic measurements are employed to investigate the allometric scaling relationships between city size and urban area. One is urban population, and the other, the land-use area. The city population is estimated by using census data, and it is greater than the traditional non-agricultural population in China; the area of urban construction land is evaluated from the current land-use maps using technology of geographical information system (GIS), and it is close to the surface area of built district. The data were obtained and processed by the experts/specialists from China Academy of Urban Planning and Design (CAUPD). Compared with the statistical data from yearbooks of Ministry of Housing and Urban-Rural Development of the People's Republic of China (MOHURD), the data quality is good and satisfying. Especially, the geographical scope of population distribution corresponds to the urban area in the main (Table 1). In fact, the quality of a dataset can be reflected by modeling process. For a city, the randomly fabricative data have no trend and no regularity, the deliberately fabricative data have trend but no regularity, and the properly experimental and observational data have both trend and regularity.

Table 1 The datasets of urban population and area of urban construction land of Xinyang city

| Year($n$) | 1949 | 1979 | 1993 | 1999 | 2004 | 2013 |
|---|---|---|---|---|---|---|
| Time ($t=n$-1949) | 0 | 30 | 44 | 50 | 55 | 64 |
| City population ($P_t$) [10,000] | 3.7 | 15.0 | 33.0 | 32.0 | 46.0 | 67.0 |
| Urban area ($A_t$)[km$^2$] | 4.50 | 17.15 | 25.10 | 31.63 | 42.89 | 89.11 |

## 3.2 Results

On the whole, the urban growth of Xinyang followed the scaling law of allometry. The significance level of statistical inference is $\alpha=0.05$. For urban area and population, the model of allometric growth, equation (1), can be rewritten as

$$A_t = aP_t^b, \tag{15}$$

where $a$ denotes the proportionality coefficient, and $b$ refers to the scaling exponent. The observational data from 1949 to 2004 can be well fitted to the power function (Figure 1). By the least squares computation, a model of longitudinal allometry can be built as follows

$$\hat{A}_t = 1.5038 P_t^{0.8605}, \tag{16}$$

where $P_t$ refers to urban population, and $A_t$ to urban construction land, the hat "^" indicates estimated value or predicted value. The goodness of fit is about $R^2=0.9817$. The scaling exponent is around $b=0.8605$, which is close to the theoretical expectation 0.85 (Chen, 2010). However, the data point of 2013 seems to be an outlier, which will be discussed in the next section. If we fit the data points from 1949 to 2013 in Table 1 to equation (15), the model will be as below:

$$\hat{A}_t = 1.2241 P_t^{0.9498}. \tag{17}$$

The squared correlation coefficient is around $R^2=0.9618$, and the scaling exponent is about $b=0.9498$. Compared with the population size of Xinyang in 2013, the urban area is too large. In other word, the city population and urban area did not match with each other in recent year.

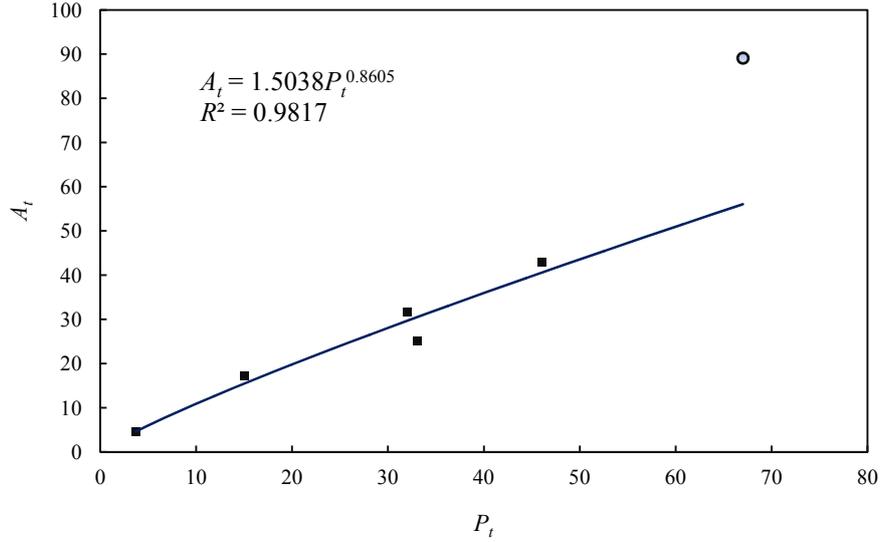

**Figure 1 The allometric relationship between the urban population and area of urban construction land of Xinyang city (1949-2004/2013)**

[**Note:** Because of scaling breaking, the observational data point in 2013, a dot, is treated as an exceptional value and not taken into account in the least squares calculation of the scaling exponent.]

If the relationship between the urban area and population of a city followed the allometric scaling law, the city population and land-use area would take on the exponential growth. In fact, the city population of Xinyang from 1949 to 2013 can be fitted into the exponential function, and the result is

$$\hat{P}_t = P_0 e^{r_p t} = 3.8208 e^{0.0451 t}, \tag{18}$$

where $P_0$ refers to the initial value of city population, and $r_p$ to the relative rate of population growth. The determination coefficient is about $R^2=0.9911$ (Figure 2a). Fitting the data points of urban construction land from 1949 to 2004 to the exponential function yields

$$\hat{A}_t = A_0 e^{r_a t} = 4.6599 e^{0.0396 t}, \tag{19}$$

where $A_0$ refers to the initial value of urban area, and $r_a$ to the relative rate of area growth. The goodness of fit is about $R^2=0.9923$. Where urban area is concerned, the data point in 2013 is an outlier (Figure 2b). If we fit the observational values of urban construction land from 1949 to 2013 into the exponential function, the squared correlation coefficient will descend from 0.9923 to 0.9738.

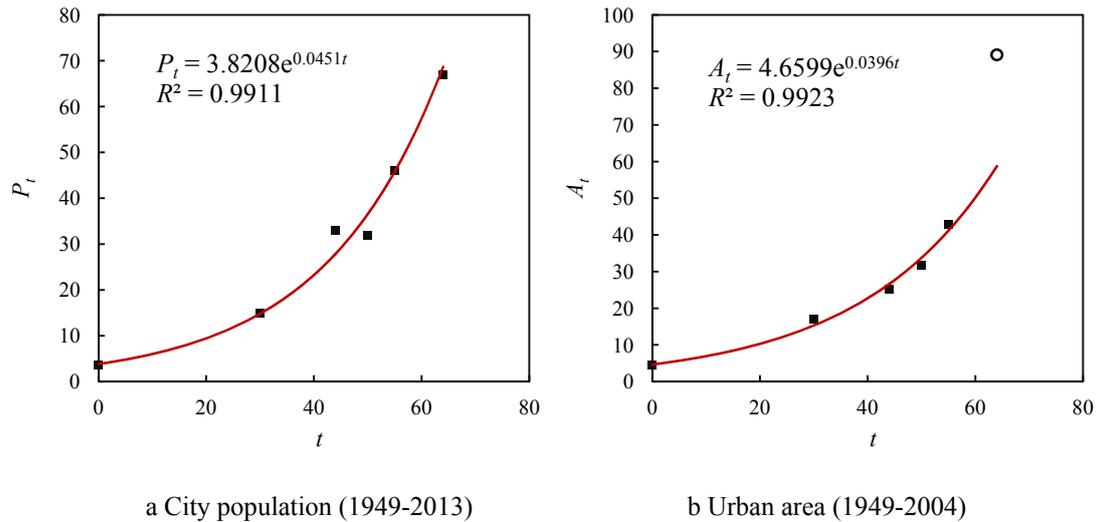

a City population (1949-2013)  b Urban area (1949-2004)

**Figure 2 The patterns of exponential growth of the urban population and area of urban construction land of Xinyang city (1949-2004/2013)**

[**Note:** The observational data point in 2013, a small circle, is regarded as an exceptional value and not taken into account in the least squares computation.]

The exponential function suggests an unlimited growth, which indicates a positive feedback process. In the real world, nothing can increase in size at a constant relative rate of growth because of the constraints of spatio-temporal conditions or the limitation of matter, energy, and information. Therefore, the model of exponential growth is valid for a short term, especially for an initial stage. In other words, there is a certain fitting scale in a growth curve for the exponential function. Owing to the limits of geographical space, natural resources, attraction, infrastructure, and so on, there must be a capacity for a city's development. Thus, the model of exponential growth should be substituted with the model of logistic growth, which has a capacity parameter indicative of the limits to urban development. By the least squares based on a logarithmic linear relation and data points of Xinyang from 1949 to 2013, we can build a logistic growth model of urban population as follows

$$\hat{P}_t = \frac{P_{\max}}{1+(P_{\max}/P_0-1)e^{-k_p t}} = \frac{487}{1+128.6587e^{-0.0471t}}, \quad (20)$$

where $P_0$ is the initial value of city population $P_t$, $P_{\max}$ is the parameter of population capacity, i.e., the maximum value of population size $P_t$, $k_p$ denotes the original rate of growth of city population $P_t$. The goodness of fit is about $R^2=0.9914$ (Figure 3). Based on the observational values from 1949 to 2004, a logistic growth of urban area of Xinyang can be made as below:

$$\hat{A}_t = \frac{A_{\max}}{1+(A_{\max}/A_0-1)e^{-k_a t}} = \frac{435}{1+92.9905e^{-0.0411t}}, \quad (21)$$

where $A_0$ is the initial value of urban area $A_t$, $A_{\max}$ is the maximum value of $A_t$, and $k_p$ refers to the original rate of growth of $A_t$. The goodness of fit is around $R^2=0.9925$ (Figure 4). The data point for 2013 is treated as exceptional value and not taken into consideration for modeling process.

Table 2 The predicted value of the city population size and the area of urban construction land of Xinyang city (1949-2014)

| Year $n$ | Time $t$ | Population $P_t$ | Urban area $A_t$ | Year $n$ | Time $t$ | Population $P_t$ | Urban area $A_t$ |
|---|---|---|---|---|---|---|---|
| 1949 | 0 | 3.7560 | 4.6281 | 1982 | 33 | 17.2852 | 17.4307 |
| 1950 | 1 | 3.9358 | 4.8202 | 1983 | 34 | 18.0882 | 18.1315 |
| 1951 | 2 | 4.1240 | 5.0201 | 1984 | 35 | 18.9269 | 18.8593 |
| 1952 | 3 | 4.3212 | 5.2281 | 1985 | 36 | 19.8029 | 19.6149 |
| 1953 | 4 | 4.5278 | 5.4447 | 1986 | 37 | 20.7177 | 20.3992 |
| 1954 | 5 | 4.7441 | 5.6702 | 1987 | 38 | 21.6727 | 21.2134 |
| 1955 | 6 | 4.9706 | 5.9049 | 1988 | 39 | 22.6697 | 22.0582 |
| 1956 | 7 | 5.2079 | 6.1491 | 1989 | 40 | 23.7101 | 22.9349 |
| 1957 | 8 | 5.4563 | 6.4033 | 1990 | 41 | 24.7958 | 23.8444 |
| 1958 | 9 | 5.7165 | 6.6678 | 1991 | 42 | 25.9284 | 24.7878 |
| 1959 | 10 | 5.9889 | 6.9431 | 1992 | 43 | 27.1097 | 25.7662 |
| 1960 | 11 | 6.2741 | 7.2296 | 1993 | 44 | 28.3415 | 26.7807 |
| 1961 | 12 | 6.5727 | 7.5277 | 1994 | 45 | 29.6257 | 27.8324 |
| 1962 | 13 | 6.8853 | 7.8378 | 1995 | 46 | 30.9641 | 28.9225 |
| 1963 | 14 | 7.2126 | 8.1605 | 1996 | 47 | 32.3587 | 30.0521 |
| 1964 | 15 | 7.5552 | 8.4962 | 1997 | 48 | 33.8115 | 31.2225 |
| 1965 | 16 | 7.9138 | 8.8454 | 1998 | 49 | 35.3244 | 32.4348 |
| 1966 | 17 | 8.2891 | 9.2086 | 1999 | 50 | 36.8995 | 33.6902 |
| 1967 | 18 | 8.6818 | 9.5865 | 2000 | 51 | 38.5389 | 34.9900 |
| 1968 | 19 | 9.0929 | 9.9795 | 2001 | 52 | 40.2445 | 36.3354 |
| 1969 | 20 | 9.5230 | 10.3882 | 2002 | 53 | 42.0186 | 37.7277 |
| 1970 | 21 | 9.9730 | 10.8132 | 2003 | 54 | 43.8633 | 39.1680 |

| 1971 | 22 | 10.4439 | 11.2551 | 2004 | 55 | 45.7805 | 40.6577 |
| 1972 | 23 | 10.9364 | 11.7146 | 2005 | 56 | 47.7726 | 42.1981 |
| 1973 | 24 | 11.4516 | 12.1923 | 2006 | 57 | 49.8415 | 43.7903 |
| 1974 | 25 | 11.9905 | 12.6889 | 2007 | 58 | 51.9894 | 45.4357 |
| 1975 | 26 | 12.5541 | 13.2051 | 2008 | 59 | 54.2184 | 47.1354 |
| 1976 | 27 | 13.1434 | 13.7417 | 2009 | 60 | 56.5306 | 48.8907 |
| 1977 | 28 | 13.7596 | 14.2993 | 2010 | 61 | 58.9279 | 50.7029 |
| 1978 | 29 | 14.4038 | 14.8787 | 2011 | 62 | 61.4124 | 52.5731 |
| 1979 | 30 | 15.0772 | 15.4807 | 2012 | 63 | 63.9860 | 54.5024 |
| 1980 | 31 | 15.7811 | 16.1062 | 2013 | 64 | 66.6505 | 56.4922 |
| 1981 | 32 | 16.5166 | 16.7559 | 2014 | 65 | 69.4078 | 58.5433 |

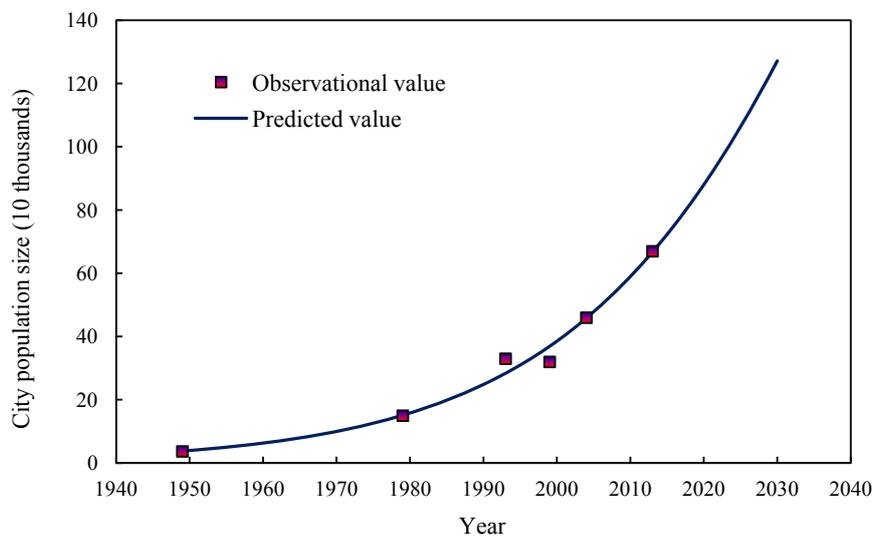

**Figure 3 The logistic growth curve of the urban population of Xinyang (1949-2030)**

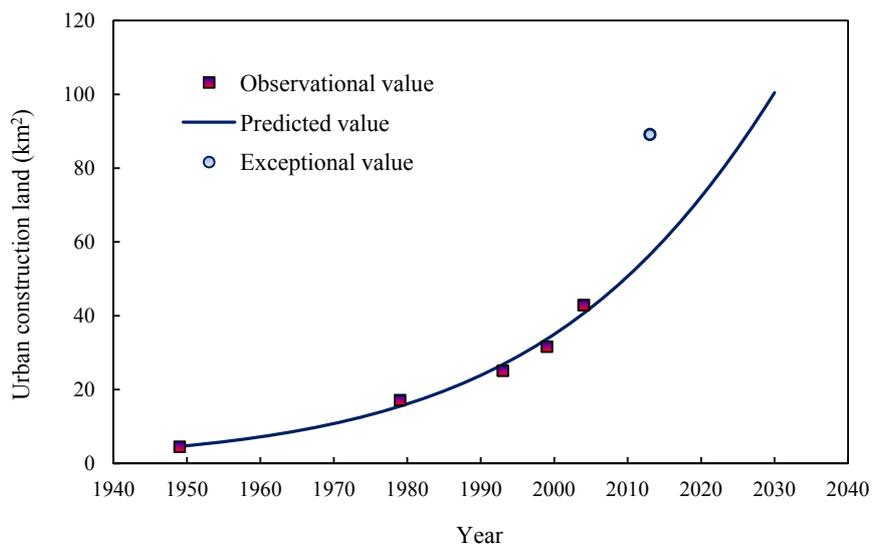

**Figure 4 The logistic growth curves of the area of urban construction land of Xinyang**

**(1949-2030)**

[**Note:** The observational data point in 2013, a dot, is treated as an exceptional value and not taken into account in the least squares calculation.]

Now, we can validate the inference that an allometric scaling relation can be approximately derived from two logistic functions for the initial stage if the time span is not large. Using equations (19) and (20), we can generate the predicted values of urban area and population (Table 2). Fitting the predicted values from 1949 to 2013 to the equation of allometric growth yields

$$\hat{A}_t = 1.4648 P_t^{0.8692}, \tag{22}$$

which is similar to equation (16). The goodness of fit is about $R^2=0.999999$ (Figure 5). The value of the scaling exponent, 0.8692, is close to the ratio of $k_a$ to $k_p$, 0.8723. This suggests that, for a short-term growth corresponding to the initial stage of urban development, the logistic models do support the allometric scaling relation. In other words, equation (1) can be derived from equations (7) and (8) when $A_t<A_{max}/2$, $P_t<P_{max}/2$, where $A_{max}$ and $P_{max}$ denote the capacity values of urban area and population.

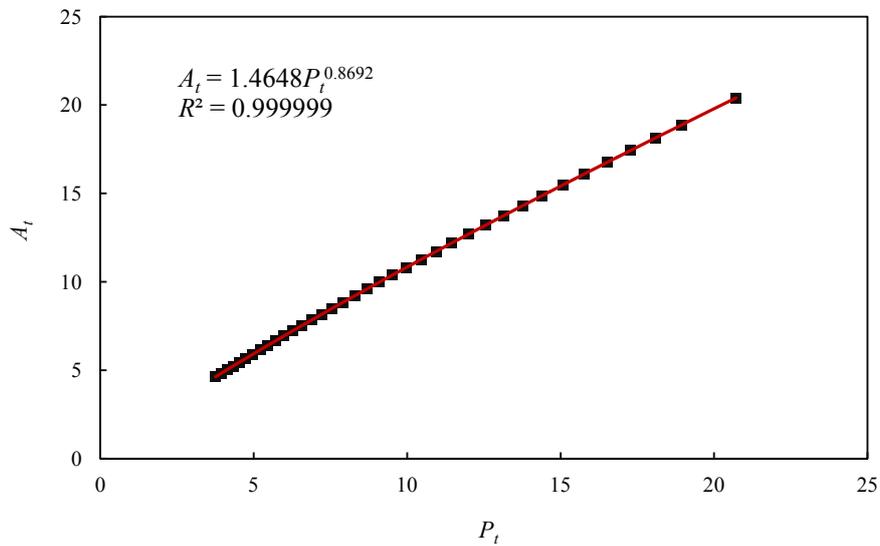

**Figure 5 The allometric relation between the urban population and the area of urban construction land of Xinyang city based on logistic growth (1949-2013)**

However, if a city is large enough, and we consider a long course of city development that follows logistic rule, the thing will be different. That is, if the stage with area $A_t>A_{max}/2$ and $P_t>P_{max}/2$ is taken into account, the simple allometric scaling relationship between urban area and population will not be supported by observational values. In this instance, we need the allometric scaling equation based on logistic growth. Define a replacement quotient of urban area $U_t = A_t/(A_{max}-A_t)$ and a replacement quotient of city population $V_t = P_t/(P_{max}-P_t)$. Then equation (14) can be rewritten as

$$U_t = \eta V_t^\sigma, \tag{23}$$

or

$$\frac{A_t}{A_{max} - A_t} = \eta \left(\frac{P_t}{P_{max} - P_t}\right)^\sigma, \tag{24}$$

in which the parameters can be expressed as $\sigma = k_a/k_p$, $\eta = A_0/(A_{max}-A_0)/[P_0/(P_{max}-P_0)]^\sigma$. Using the least squares calculation, we can fit the data in Table 1 to equation (24), and the model for years from 1949 to 2004 is

$$\frac{\hat{A}_t}{435 - \hat{A}_t} = 0.7249 \left(\frac{P_t}{487 - P_t}\right)^{0.8637}. \tag{25}$$

The determination coefficient is about $R^2=0.9808$, and the scaling exponent is around $\sigma=0.8637$ (Figure 6). Further, fitting the data in Table 2 to equation (24) yields

$$\frac{\hat{A}_t}{435 - \hat{A}_t} = 1.3441 \left(\frac{\hat{P}_t}{487 - \hat{P}_t}\right)^{0.8723}. \tag{26}$$

This is a perfect fit and the goodness of fit is 1 because the urban area and city population are predicted values rather than observational values. The scaling exponent is $\sigma=k_a/k_p \approx 0.8723$, where $k_a \approx 0.0411$, $k_p \approx 0.0471$.

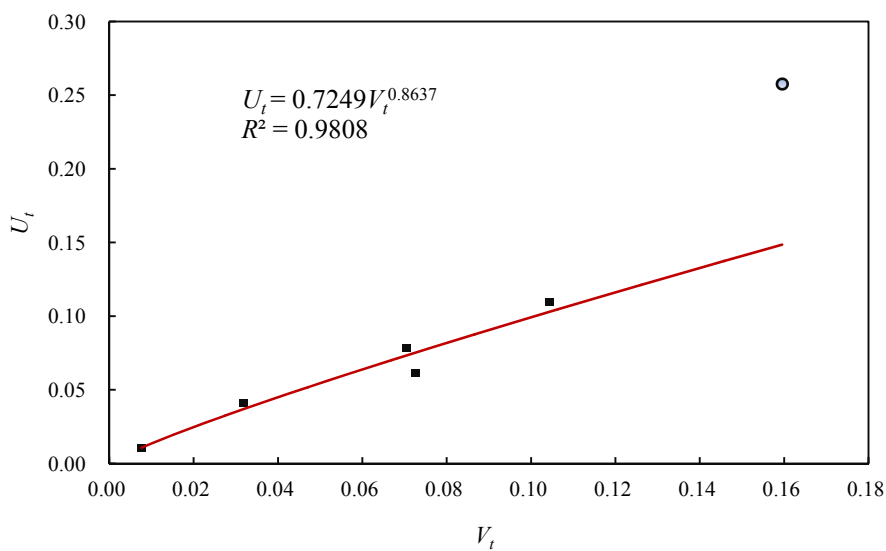

**Figure 6 The allometric relation between the replacement quotient of urban population and that of urban construction land of Xinyang city (1949-2004/2013)**

[**Note:** The observational data point in 2013, a dot, is treated as an exceptional value and not taken into account in the least squares calculation of the scaling exponent.]

# 4 Questions and discussion

For convenience of discussion, the process of logistic growth should be divided into different stages. In fact, urban growth is a process of urbanization (Knox and Marston, 2009). The level of urbanization can be measured with the ratio of urban population to total population within a geographical region, $L_t=u_t/(u_t+r_t)$, where $L_t$ refers to the urbanization level, and $u_t$ and $r_t$ to urban and rural population in a region, respectively. In many cases, the level of urbanization can be described with a logistic function. Thus the growing curve of urbanization level used to be divided into three stages: initial stage, acceleration stage, and terminal stage (Northam, 1979). However, according to the mathematical property of the logistic function, Northam's scheme of stage division is not accurate and the acceleration stage should be renamed for celerity stage, which can be further separated into acceleration stage and deceleration stage (Chen and Zhou, 2005). In other words, the logistic process of urbanization can be partitioned into four stages: initial stage, acceleration stage, deceleration stage, and terminal stage (Figure 7, Table 3). The four-stage division can be generalized to any logistic growth, and a logistic curve of a city's development falls into four successive phases.

**Table 3 The division results of growth stages of a logistic process for urbanization**

| Urbanization stage | Original classification | Revised classification | |
|---|---|---|---|
| | Three-stage division | Three-stage division | Four-stage division |
| Urban minority ($L_t<0.5$) | Initial stage | Initial stage | Initial stage |
| | Acceleration stage | Celerity stage | Acceleration stage |
| Urban majority ($L_t>0.5$) | | | Deceleration stage |
| | Terminal stage | Terminal stage | Terminal stage |

**Note**: The original classification comes from Northam (1979).

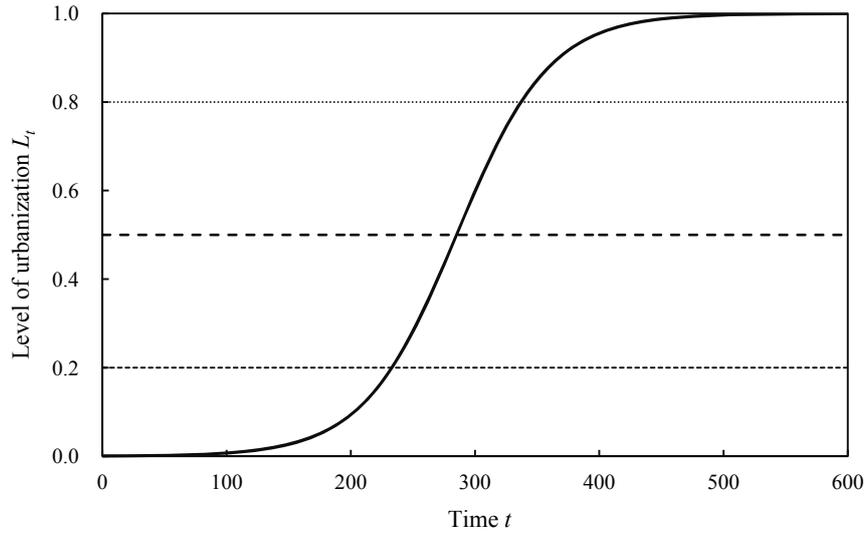

**Figure 7 A logistic curve of an urbanization process divided into four growth stages**

**Note**: The capacity parameter of urbanization level is $L_{max}=1$, so the inflexion of the logistic growth is at 1/2. Otherwise, the inflexion will be at $L_{max}/2$ instead of 1/2.

Now, we can see that the allometric scaling relation based on exponential growth is different from that based on logistic growth. If a city grows in the exponential way, the global relationship between the area measurement and population size will follow a power law, equation (15), which is called allometric scaling law. However, if the city grows in the logistic way rather exponential way, the global scaling relation between urban area and population size will break. In this case, only the initial stage of growth satisfies the allometric scaling relation between urban area and population. This is a kind of local scaling relation of allometric growth. For the global course of urban growth, the scaling relation comes between the replacement quotient of urban area, $U_t=A_t/(A_{max}-A_t)$, and that of city population, $V_t=P_t/(P_{max}-P_t)$. The process of urbanization is in fact a dynamics of spatial replacement: nonurban population is replaced by urban population, and nonurban land is replaced by urban land (Chen, 2012).

To realize the similarities and differences between the allometric scaling based on exponential growth and that based on logistic growth, a comparison between the two can be drawn as follows (Table 4). For the mathematical models, the allometric scaling based on exponential growth can be expressed with equation (15), while the scaling relation based on logistic growth can be described by equation (23) or equation (24). For the former, the scaling exponent is $b=r_a/r_p$; for the latter, the scaling exponent is $\sigma=k_a/k_p$. As for Xinyang from 1949 to 2004/2013, we have $b\approx 0.0396/0.0451\approx$ 0.8770 and $\sigma \approx 0.0411/0.0471 \approx 0.8723$. The scaling exponent values fall between 2/3 and 1 (Chen, 2010; Lee, 1989). The measurements related to the scaling exponents are the ratio of the relative rate of area growth to that of population growth. For the exponential growth, the relative rate of growth of urban area and that of city population are all constants: $r_a = dA_t/(A_t dt)$, $r_p = dP_t/(P_t dt)$. Thus we have a ratio such as

$$s = \frac{r_a}{r_p} = \frac{dA_t/(A_t dt)}{dA_t/(A_t dt)} = b, \qquad (27)$$

where $s$ denotes the ratio of the relative rate of growth of urban area to that of city population. However, for the logistic growth, the relative rate of growth of urban area and that of city population are not constant. The constants are the initial rates of growth $k_a = dA_t/(A_t dt)/(1-A_t/A_{max})$ and $r_p = dP_t/(P_t dt)/(1-P_t/P_{max})$. The ratio of the relative growth rates can be theoretically expressed as

$$s = \frac{dA_t}{A_t dt} / \frac{dP_t}{P_t dt} = \frac{k_a}{k_p} \frac{1 - A_t/A_{max}}{1 - P_t/P_{max}} \neq \sigma. \qquad (28)$$

This is a variable instead of a constant. For the discrete data, the ratio can be rewritten as

$$\hat{s} = \frac{A_t - A_{t-1}}{A_{t-1}} / \frac{P_t - P_{t-1}}{P_{t-1}} \rightarrow \frac{k_a}{k_p} \frac{1 - A_t/A_{max}}{1 - P_t/P_{max}} = s, \qquad (29)$$

where the arrow "→" indicates "be close to". For the relationship between urban area and population, the ratio $s$ is always mistaken as the scaling exponent $b$, which is always close to $\sigma$. At the first stage, the $s$ value is near a constant $\sigma$ because $A_t/A_{max}$ and $P_t/P_{max}$ are close to zero. However, if $A_t > A_{max}/2$ and $P_t > P_{max}/2$, the $s$ value will go up and up (Figure 8).

**Table 4 A comparison between the allometric scaling model based on exponential growth and that based on logistic growth**

| Item | Exponential allometry | Logistic allometry |
|---|---|---|
| Description | A simple power-law relation based on exponential growth | A complex power-law relation based on logistic growth |
| Model | $A_t = aP_t^b$ | $\dfrac{A_t}{A_{max} - A_t} = \eta(\dfrac{P_t}{P_{max} - P_t})^\sigma$ |

| Scaling exponent | $b = \dfrac{r_a}{r_p}$ | $\sigma = \dfrac{k_a}{k_p}$ |
|---|---|---|
| Rate of growth | $r_a = \dfrac{dA_t}{A_t dt}, r_p = \dfrac{dP_t}{P_t dt}$ | $k_a = \dfrac{dA_t}{A_t dt}/(1-\dfrac{A_t}{A_{max}}), k_p = \dfrac{dP_t}{P_t dt}/(1-\dfrac{P_t}{P_{max}})$ |
| Ratio of growth rates | $s = \dfrac{r_a}{r_p} = b$ | $s = \dfrac{k_a}{k_p}\dfrac{1-A_t/A_{max}}{1-P_t/P_{max}} \neq \sigma$ |
| Per capita urban area | $L_t = \dfrac{A_0}{P_0}e^{(r_a-r_p)t}$ | $L_t = \dfrac{A_{max}}{P_{max}}\dfrac{1+(P_{max}/P_0-1)e^{-k_p t}}{1+(A_{max}/A_0-1)e^{-k_a t}}$ |

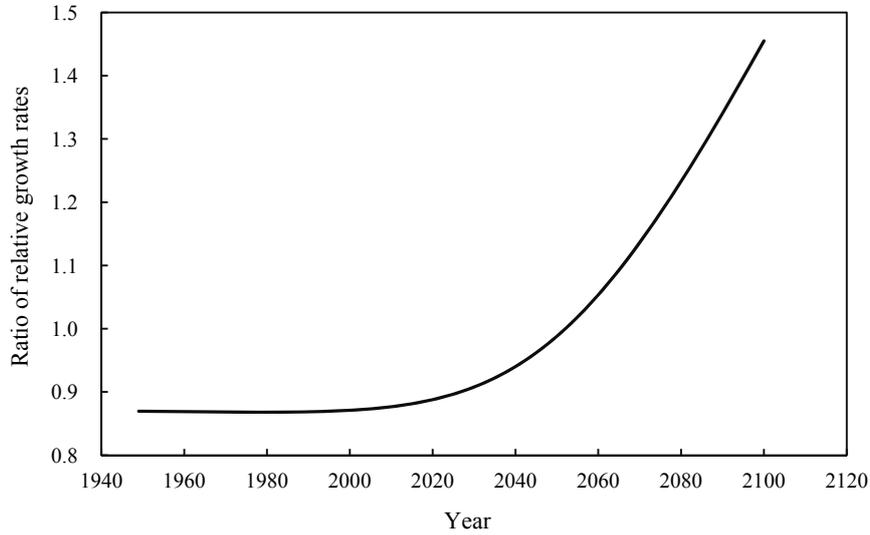

**Figure 8 The ratio of the relative rate of growth of the urban area to that of the city population of Xinyang (1949-2100)**

The scaling exponent is associated with the per capita area of urban construction land. According to equation (15), the per capita urban area based on exponential growth is

$$L(P_t) = \dfrac{A_t}{P_t} = aP_t^{b-1}, \qquad (30)$$

where $L(P)$ denotes the per capita area changing with population size. According to equations (18) and (19), the per capita urban area based on exponential growth is

$$L_t = \dfrac{A_0}{P_0}e^{(r_a-r_p)t}, \qquad (31)$$

in which $L_t$ denotes the per capita area changing over time. As for Xinyang, two models are as follows

$$\hat{L}(P_t) = \frac{\hat{A}_t}{P_t} = 1.5038 P_t^{-0.1395}, \tag{32}$$

$$\hat{L}_t = 1.2196 e^{-0.0056t}. \tag{33}$$

However, based on logistic growth, the changing curves of per capita urban area is different. According to equations (20) and (21), the per capita urban area over time is as below

$$L_t = \frac{A_t}{P_t} = \frac{A_{max}}{P_{max}} \frac{1+(P_{max}/P_0-1)e^{-k_p t}}{1+(A_{max}/A_0-1)e^{-k_a t}}. \tag{34}$$

As for Xinyang, we have

$$\hat{L}_t = \frac{\hat{A}_t}{\hat{P}_t} = 0.8932 \frac{1+128.6587 e^{-0.0471t}}{1+92.9905 e^{-0.0411t}}, \tag{35}$$

by which the changing curve of per capita urban area over time looks like a scoop (Figure 9).

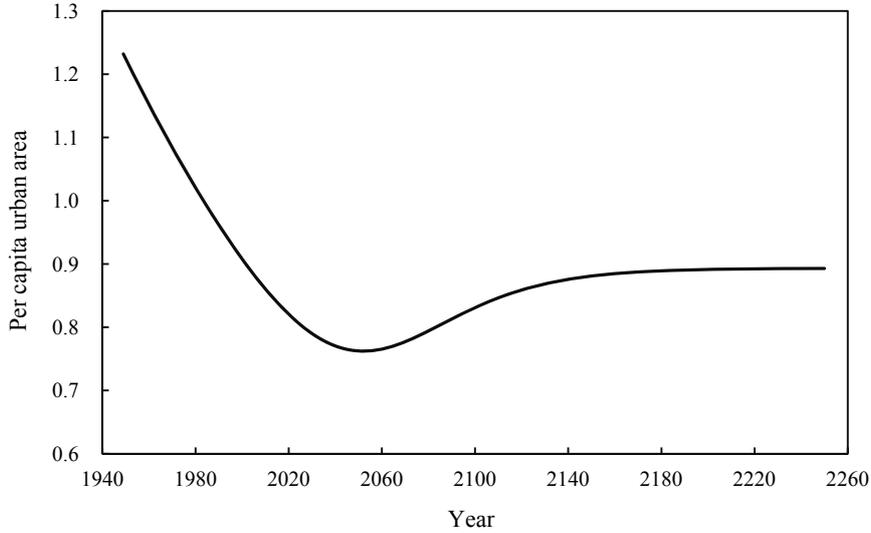

**Figure 9 The per capita area of urban construction land of Xinyang over time (1949-2250)**

The rank-size patterns of cities show a fact the population growth goes before land expansion. In log-log plots, the scaling range of population size distribution is significantly wider than that of area size distribution of cities (Chen and Zhou, 2008). The urban growth of Xinyang city from 1949 to 2004 lent support to this suggestion. Based on equations (20) and (21), the curves of the absolute rates of growth of urban population and area take on two wave crests which do not keep chime with one another. The peak of population growth rate is ahead of that of land-use expansion (Figure 10). However, this dynamical process of the city was broken down since 2006. From then on, urban land expansion has been going before population increase.

Now, the outlier of Xinyang's urban growth can be clarified by observational facts. As stated above, the urban area in 2013 is an exceptional value, which departs from the trend line of logistic growth. This results in an abnormal behavior of urban allometric growth. The urban area and population of Xinyang follow the allometric scaling law from 1949 to 2004, and the scaling exponent is about $b \approx 0.8605$. However, during the period between 2007 and 2013, the scaling relation was interrupted and the scaling exponent suddenly went up near $b \approx 1.49$. This is a kind of scaling breaking. The cause is what is called "city-making movement" in the background of bubble economy based on realty industry expansion and fast urbanization. In fact, in 2006, a new leader of Xinyang city went to take office. In order to create the administrative performance quickly, the local leader and the officials determined to open up a new urbanized area, which is termed "Yangshan district". Consequently, the urban land-use increase significantly surpassed the population growth. Normally, at the initial stage and acceleration stage, land expansion should lag behind population increase (Figure 10). Due to the human factors against self-organized evolution, the conventional logistic model of urban area expansion will be replaced by a quadratic logistic function as follows

$$A_t = \frac{A_{\max}}{1 + (A_{\max}/A_0 - 1)e^{-k_a t^2}}. \tag{36}$$

An inference can be drawn from the above analytical process that urban evolution in the developing countries such as China follows the allometric growth law, but the allometric scaling can be disturbed by the governmental behaviors based on command economy. The subjective behavior of the officials in local governments may change the mode of urban growth.

The logistic allometry can be generalized to model the process of urbanization of a country. As a matter of fact, the allometric analysis is originally employed to investigate the relationships between urban and rural population of an urbanized region (Naroll and Bertalanffy, 1956). However, the exponential allometry is only suitable for the initial stage of urbanization. For a long period of urbanization, the simple model of allometric scaling is not convincing. For example, for the United States of America, the initial stage (1790-1830) and acceleration stages (1840-1900) can be described with an allometric relation such as

$$u(t) = 0.0000000109 r(t)^{2.0001}, \quad (38)$$

where $u(t)$ refers to the urban population of time $t$, and $r(t)$ to the rural population of the same time. The goodness of fit is about $R^2=0.9842$. This is an allometry based on the exponential growth of urban and rural population. It is not valid for the full course of urbanization from 1790 to 2010. For the initial stage (1790-1830), acceleration stage (1840-1900), deceleration stage (1910-1990), and early terminal stage (2000-2010), the allometric scaling is as below:

$$\frac{u(t)}{273300000 - u(t)} = 0.0388 \left[ \frac{r(t)}{63135000 - r(t)} \right]^{1.4269}, \quad (39)$$

which is an allometry based on the logistic growth of the U.S. urban and rural population. The coefficient of determination is around $R^2=0.9522$. The capacity values of urban and rural population are estimated as $u_{max}=273300000$ and $r_{max}=63135000$, respectively. In fact, strictly speaking, the logistic allometry is not yet applicative for the terminal stage of American urbanization. Equation (39) is an approximate relation, and we need a new model for the allometric modeling of urbanization.

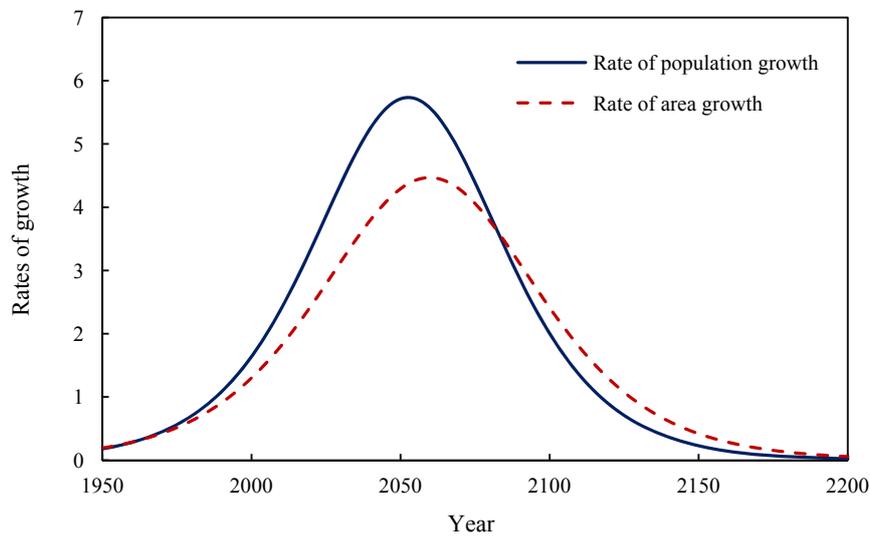

**Figure 10 The unimodal curves of the growth rates of urban population and area of Xinyang (1949-2200)**

# 5 Conclusions

An allometric scaling can be divided into exponential allometry and logistic allometry. The former is based on exponential growth, while the latter is based on logistic growth. In this paper, a theoretical model of logistic allometry is proposed to describe the global process of urban evolution, and the exponential allometry is approximately suitable for the initial phase of logistic growth. The new allometric scaling can be empirically demonstrated with the observational datasets of urban area and population. The main points of this article are as follows.

**First, there is an allometric scaling relation based on logistic growth between urban area and population.** The logistic allometry is for spatial replacement quotients rather than the simple basic measurements such as urban area and population. Different from the exponential allometry, the scaling exponent of the logistic allometry is not equal to the ratio of relative rate of growth of one element to that of another element. The ratio of relative growth rates is a variable instead of a constant. For the initial stage or even the acceleration stage, the ratio of relative growth rates is less than 1; however, for the deceleration stage and the terminal stage, the ratio of relative growth rates will be greater than 2, or even go up and up.

**Second, the exponential allometry can be regarded as an approximation of the logistic allometry of urban evolution.** The logistic allometry is more universal than the exponential allometry. The exponential allometric scaling is usually a local concept and can be used to describe the initial stage of city development. The logistic allometric scaling is a global concept and can be employed to characterize the whole process of urban growth, including initial stage, acceleration stage, deceleration stage, and terminal stage. For a young city, we can use the exponential allometry to model its growth; however, for a ripe city, the logistic allometry should be employed to reveal its dynamic process in the course of growth.

**Third, the model of logistic allometry can be generalized to describe other dynamic process of urban evolution such as urbanization.** An urbanization process is a logistic process or a generalized logistic process. The exponential allometry used to be employed to research urbanization. However, only at the initial stage, urbanization follows the simple allometric scaling law. The growth curve of urban population in a region can often be described with logistic function rather than exponential function, and at the initial, acceleration, and deceleration stages,

the rural population growth can also be modeled with logistic function. As a result, urbanization process follows the law of logistic allometry instead of exponential allometry.

## Acknowledgement

This research was sponsored by the National Natural Science Foundation of China (Grant No. 41171129). The supports are gratefully acknowledged. The author would like to thank Dr. Li Zhang of China Academy of Urban Planning and Design for providing the essential data on Xinyang's urban population and construction land area.